\documentclass{ifacconf}

\usepackage{graphicx}      
\usepackage{natbib}        
\usepackage{amsmath}
\usepackage{amsfonts}
\usepackage{amssymb}
\usepackage{mathtools}
\usepackage{xcolor}

\begin{document}
\begin{frontmatter}

\title{Nonlinear Controllability Assessment of Aerial Manipulator Systems using Lagrangian Reduction \thanksref{footnoteinfo}} 

\thanks[footnoteinfo]{This work was funded in part by a contract from the NASA Jet Propulsion Laboratory, and by a NASA NSTRF fellowship for the second author.}

\author[First]{Skylar X. Wei} 
\author[First]{Matthew R. Burkhardt} 
\author[First]{Joel Burdick}

\address[First]{California Institute of Technology, 
   Pasadena, CA 91125 USA (e-mail: swei@caltech.edu; burkhardt.matt.r@gmail.com; jwb@robby.caltech.edu).}

\begin{abstract}                
This paper analyzes the nonlinear Small-Time Local Controllability (STLC) of a class of underatuated aerial manipulator robots.
We apply methods of Lagrangian reduction to obtain their lowest dimensional equations of motion (EOM). The symmetry-breaking potential energy terms are resolved using advected parameters, allowing full $SE(3)$ reduction at the cost of additional advection equations. The reduced EOM highlights the shifting center of gravity due to manipulation and is readily in control-affine form, simplifying the nonlinear controllability analysis. Using Sussmann's sufficient condition, we conclude that the aerial manipulator robots are STLC near equilibrium condition, requiring Lie bracket motions up to degree three. 
\end{abstract}

\begin{keyword}
Controllability; Geometric Mechanics; Lagrangian Reduction; Flying robots; UAVs.
\end{keyword}

\end{frontmatter}

\section{Introduction}
Aerial manipulators, which combine a multi-rotor aerial platform with a multi-jointed robot arm, have the mobility of multi-rotors and can interact with the environment utilizing its manipulator. A variety of aerial manipulators have been introduced, ranging from a fully-actuated aerial base with a three-link manipulator (Sanchez-Cuevas et al.(2020)) to a helicopter with a seven degree-of-freedom manipulator (Kondak et al. (2014)). As summarized by Ruggiero et al. (2018), no matter the complexity of the fully-actuated appended arm, the underactuation of the aerial base poses limitations on the overall system ability.

Despite many successful designs, there is no formal guideline for designing an aerial manipulator based on task objectives, nor has there been any analytical assessment of the limitation inherent in an underactuated base. To the best of the authors' knowledge, there has not been any theoretical work studying the underactuated aerial system's controllability properties. Controllability, which describes a system's ability to drive its states to arbitrary values by choices of feasible inputs, is widely used in the study of stabilization, feedback controller design, and state-space reduction. Nonlinear controllability analysis is more challenging than the linear case, but it must be considered for aerial manipulators to ensure sufficient manipulability.

To tackle the algebraic complexity of nonlinear controllability analysis, we establish a lower-dimensional EOM by utilizing Lagrangian reduction.  This reduction technique yields a set of the first-order EOM on a lower-dimensional phase space that excludes the symmetry group, which is algebraically simpler in the subsequent controllability assessment. Methods of Lagrangian reduction have successfully been applied to evaluating controllability of complex undulatory robots (\cite{Ostrowski_1996}) and the modeling and control of spherical robotic vehicles (\cite{Burkhardt_2018}). We consult our earlier work (\cite{Burkhardt_advectedparam}) to tackle the symmetry-breaking potential terms of aerial manipulators via advected parameters.

We analyze nonlinear controllability under small-time, local, and near-equilibrium conditions, using the sufficient conditions given by \cite{Sussmann1987AGT}, which applies to systems with a control-affine form, a non-zero drift, and bilateral (both positive and negative) control inputs. However, aerial manipulators have unilateral control inputs (non-negative thrust), making them ill-suited for a direct application. Based on the rotor dynamics, we select bilateral rotor RPM rate as the replacement control inputs. Our controllability result can be viewed as a certification of an aerial manipulator's fundamental capability to control the multirotor platform during aerial manipulation. The set of reduced EOM is mathematically compact and reveals geometric properties that could serve in the motion planning and controller synthesis process.

\textbf{Paper Outline.} Section \ref{ses:sys_des} defines the class of aerial manipulator system studied. Section \ref{ses:prel} reviews the basics of reduction and reconstruction under broken symmetries, as well as controllability of nonlinear systems. The EOM for a general class of aerial manipulator system is developed in Section \ref{ses:sys_dyn}. Section \ref{ses:nonlinearC} presents the main contribution: proofs that our class of aerial manipulators are Small-Time Locally Accessible and Small-Time Locally Controllable. We conclude in Section \ref{ses:Conclusion} with future applications.

\section{Systems Description}
\begin{figure}
    \centering
    \includegraphics[width=8.4cm]{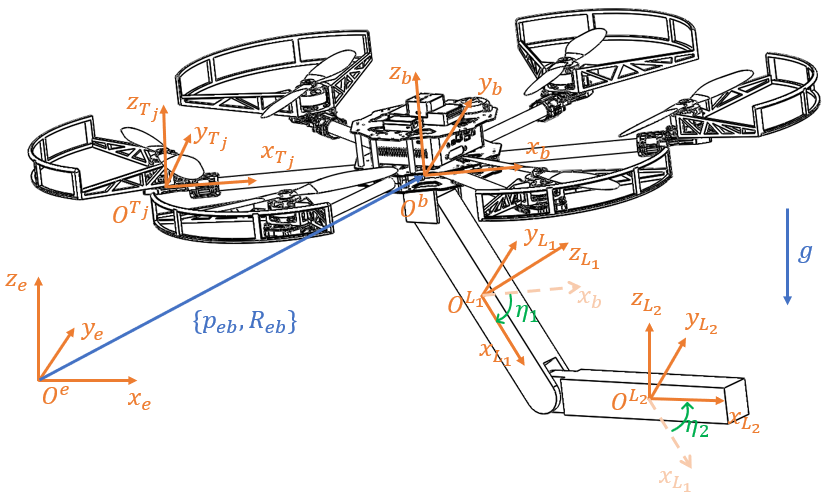}
    \caption{Pictorial illustration of relevant reference frames for the defined aerial manipulator system.}
    \label{fig:coordinate_frames}
\end{figure}

We consider the following  class of Aerial Manipulators:

\textbf{Definition:}
A multirotor aerial platform with the following characteristics is an \textbf{Aerial Manipulator (AM)}:
\begin{itemize}
    \item The multi-rotor includes $n$-pairs of identical rotors where $n\geq 2$. Each rotor pair consists of one clockwise (CW) and one counterclockwise (CCW) rotating rotor, distributed in a cross configuration.  
    \item All rotor thrusts point in the positive direction of the $z_b$ axis, and the distances from their rotating axis to the $z_b$ axis are equal, see Fig. \ref{fig:coordinate_frames}.
    \item The 2-link manipulator is attached to the multi-rotor's geometric center and operates in $x_b-z_b$ plane. Each link is approximated by a uniform cylinder.
    \item All system components are rigid and complex fluid structure interactions are ignored.
\end{itemize}
The manipulator is limited to  planar operation only for simplicity: it is sufficiently complicated to demonstrate the effectiveness of the reduction process and the dynamic coupling between the multi-rotor and arm dynamics. 

Our model is  derived using the following reference frames:
\begin{itemize}
    \item The earth-fixed inertial frame $E=\{O^e,x_e,y_e,z_e\}$.
    \item The aerial-base body frame $B=\{O^b,x_b,y_b,z_b\}$.
    \item Manipulator $i^{th}$ link frame  $L_i=\{O^{L_i},x_{L_i},y_{L_i},z_{L_i}\}$.
    \item The $j^{th}$ rotor frame $T_j=\{O^{T_j},x_{T_j},y_{T_j},z_{T_j}\}$.
\end{itemize}
Hereafter, $s_{ab}\in \mathbb{R}^{3}$ and $R_{ab}\in SO(3)$ denote the position and orientation of the origin of frame $B$ relative to the origin of frame $A$, respectively. $\eta_1, \eta_2 \in \mathbb{S}$ be the relative joint angles. The aerial-base linear velocity in $B$ frame is defined to as $\dot{s}_{b} \triangleq R_{eb}^T\dot{s}_{eb}$. 
Using standard roll, pitch, and yaw (Euler) angles $\mathbf{\Theta}=[\phi,\theta,\psi]^T$, the angular velocity of the aerial-base in $B$ frames, $\omega_b$, is the following:
\small
\begin{equation} \label{eq:rollpitchyaw_eq}
    \omega_{b} = \left[\begin{array}{ccc}
        1 &0 &-\sin(\theta)  \\
        0 & \cos(\phi) & \sin(\phi)\cos(\theta) \\
        0 & -\sin(\phi) & \cos(\phi)\cos(\theta)
    \end{array}\right]\left[\begin{array}{c}
         \dot{\phi}  \\
         \dot{\theta} \\
         \dot{\psi}
    \end{array} \right] = T(\mathbf{\Theta})\dot{\mathbf{\Theta}}.
\end{equation}
\normalsize
Lastly, $\dot{R}_{eb} \triangleq R_{eb}S(\omega_b)$ where $S(\cdot)$ is the $3\times 3$ skew-symmetric matrix such that  $S(\omega_b)x = \omega_b \times x$.
\label{ses:sys_des}
\section{Preliminaries} \label{ses:prel}


A mechanical system is characterized by the tuple $\Sigma = (Q,L,\mathcal{T})$, where $Q$ is its finite-dimensional configuration space. $TQ$ denotes the tangent bundle of $Q$, and $T_qQ$ is the tangent space to $Q$ at $q$. $L:TQ\to \mathbb{R}$ is the system Lagrangian. $\mathcal{T}(q,\dot{q})\in T^*Q$ represents the external forces field acting on the system, where $T^*Q$ is the dual of $TQ$. 

Broadly speaking, a Lagrangian possesses a {\em symmetry} if there is an action on its arguments that renders the Lagrangian invariant. This symmetry allows the reduction of the dynamical system to a lower dimensional phase space. This paper is concerned with Lie group symmetries, which naturally arise in rigid body systems. For the rest of the paper, we will assume that $Q$ is a {\em trivial principle bundle}, i.e., $Q = G\times M$, where the fiber-space $G$ is a Lie group. Smooth manifold $M$ denotes the internal shape space. Consequently, coordinates on $q\in Q$ can be partitioned as $q=(\mathtt{g},r)$, where $\mathtt{g}\in G$ and $r\in M$. 
Associated with a Lie group is its {\em Lie algebra}, $\mathfrak{g}$, a vector space isomorphic to the tangent space at the group identity, i.e. $\mathfrak{g}\simeq T_{q}G$. 
In the context of reduction, we are interested in the {\em left action} of a Lie group $X$ on a smooth manifold $P$, which is defined as the map $\Phi_{\mathtt{x}}: P\times P\to P:p\to\mathtt{x}p$ for any $p\in P$. 
The Lie group of an aerial manipulator robot's configuration space is a semidirect product space, defined as $G = H\otimes \mathcal{V}$ where $H$ is a Lie subgroup and $\mathcal{V}$ is a vector space. We denote the Lie algebra of $H$ as $\mathfrak{h}$, and the group $H$ acts on the space $\mathcal{V}$ from the left, and on $Q$ via a left-action. The Lie algebra of the semidirect-product group can be written as $\mathfrak{g} = \mathfrak{h}\otimes T\mathcal{V} \simeq \mathfrak{h}\otimes \mathcal{V} $ with elements $(\xi_\mathtt{h},\xi_\mathcal{V}) \in \mathfrak{g}$. In local/body coordinates, $\xi_\mathtt{h} = \mathtt{h}^{-1}\dot{\mathtt{h}} \in \mathfrak{h}$ and $\xi_\mathcal{V} = \mathtt{h}^{-1}\dot{v}\in T\mathcal{V}$ where $(\dot{\mathtt{h}},\dot{v})\in TG$ is an arbitrary tangent vector. 
The {\em lifted action} results from the left-action $\Phi_\mathfrak{g
    }(q)$ on tangent vectors is defined as the map $T\Phi_\mathfrak{g}: TQ\to TQ:\, (q,v) \to (\Phi_\mathtt{g}(q),T_{q}\Phi_{\mathtt{g
    }}(v))$.
    For $\mathtt{g} = (\mathtt{h},v)$, $T_{q}\Phi_{\mathtt{g
    }}(\dot{l},\dot{p},\dot{r}) = (\mathtt{h}\dot{l},\mathtt{h}\dot{p},\dot{r})$.
    
A mechanical system possess a {\em symmetry} with respect to Lie group $G$ if its Lagrangian $L:TQ\to \mathbb{R}$ and external forces $F(q,\dot{q})$ are  Lie  Group  Invariant.
See \cite{SCHWARZ197563}.  As the appended arm moves during its manipulation tasks, the system's overall center of mass displaces, breaking symmetry in the Lagrangrian's potential. Thus, we consider the mechanics under a broken symmetry condition, which leads to a modified Euler-Poincar\'e set of equations as will be discussed next.


\subsection{Lagrangian Reduction and reconstruction}
The reduction technique based on advected parameter, introduced by \cite{HOLM19981}, is employed to reformulate the symmetry breaking potential contributions. In the context of mechanical system, an {\em advected parameter}, $\gamma(t)$, is a vector expressed in a body-fixed reference frame satisfying the following differential equation:\small
\begin{equation}\label{eq:advected_def}
    \left(\frac{d}{dt} + \mathtt{g
    }^{-1}(t) \dot{\mathtt{g
    }}(t) \right)\gamma(t)=0.
\end{equation}
\normalsize
where $ \mathtt{g}\in G$. 
For AM, we choose an advected parameter, $\gamma(t) \triangleq R_{eb}^{T}e_{3}\in \mathcal{V}$, to address the symmetry breaking gravity. Another symmetry breaking term of the AM is its dependency on the aerial base position. Let $\zeta \triangleq R_{eb}^Ts_{eb}$, together with $\gamma$, the AM potential energy can be expressed as $V(r,\gamma,\zeta)$, where $r$ denotes the shape variables.

Define the {\em Augmented Lagrangian} $\overline{L}:TG \times \mathcal{V} \to \mathbb{R}$ by augmenting the state with advected parameter $\gamma$ and base position $\zeta$. If the Augmented Lagrangian is $\Phi_{\mathtt{h}}$-invariant, then it can be reduced to $\mathfrak{h}\times \mathcal{V}\times TM$ (\cite{Burkhardt_2018}).  Applying the Lagrange-d’Alembert principle 
to the Augmented Lagrangian yields a modified form of the Euler-Poincar\'{e} equations. See \cite{Schneider_EP} for details. 
One important takeaway from Schneider's result is the following reduced variational principle:
\small
\begin{equation} \label{eq:reducedvariationalprinciple}
    \delta\left(\int_{t_0}^{t_f}K(\xi_\mathtt{h},\xi_\mathtt{\mathcal{V}},r,\dot{r})dt -\int_{t_0}^{t_f}V(r,\gamma,\zeta)\right) = 0.
\end{equation}\normalsize
Merging (\ref{eq:reducedvariationalprinciple}) with the Lagrange-d'Alembert principle yields:
\small
\begin{equation} \label{eq:gravityreduction}
    \delta\int_{t_0}^{t_f}K(\xi_\mathtt{h},\xi_\mathtt{\mathcal{V}},r,\dot{r})dt = \delta \int_{t_0}^{t_f}dV(r,\gamma,\zeta)\cdot \delta q dt,
\end{equation}\normalsize
where $-dV(r,\gamma,\zeta)$ represents conservative forces and moments arise from the potential energy. 

Based on \cite{Lewis_book}, we can conclude the dynamical equivalence between the mechanical systems $\Sigma = (Q,L,\mathcal{T})$ and the reduced system $\Sigma_r = (Q,K, -dV+\mathcal{T})$. 
At the expanse of two additional equations for $\dot{\gamma}$ and $\dot{\zeta}$, reduction by $G$ ($SE(3)$ for AMs) is made possible on an augmented configuration space.

\begin{thm} \label{thm:matt_thm}
(\cite{Burkhardt_2018}) Let a mechanical system with Lagrangian $L:=K-V$ satisfy the following:
\begin{enumerate}
    \item The configuration space $Q$ has the product structure $Q =G\times M =H\otimes\mathcal{V}\times M$, where $H$ is a Lie group, with a left action on vector space $\mathcal{V}$, and $M$ is the shape space. For AM, $H = SO(3)$, $\mathcal{V} = \mathbb{R}^{3}$, and $M$ is a smooth manifold. Hence, $(q,\dot{q}) = (h,\dot{h},\xi,\dot{\xi},r,\dot{r})$.
    \item The kinetic energy $K(q,\dot{q})$ and Augmented Lagrangian, $\overline{L} = TG\times \mathcal{V}\to \mathbb{R}$, are invariant with respect to a left $G$-action. $L(\Phi_\mathtt{g}(q),T_q\Phi_\mathtt{g}\dot{q}) = L(q,\dot{\mathtt{g}})$.
    \item The {\em Reduced Augmented Lagrangian} can be expressed as the following function form: $\overline{l}(\xi_\mathtt{h},\xi_{\mathcal{V}},\gamma,\zeta,r,\dot{r})$ where $r, \dot{r}\in M$, $\xi_\mathtt{h} \in \mathfrak{h}$, $\xi_{\mathcal{V}}\in \mathcal{V}$. $\gamma$ and $\zeta$ are the advected parameter and aerial base position.
\end{enumerate}
 The system's reduced equations of motion takes the form:
\begin{equation}\label{eq:matt_eq_1}
        \dot{\varrho} = \dot{r}^{T}\alpha(r,\gamma)\dot{r} + \dot{r}^{T}\beta(r,\gamma)\varrho +\varrho^{T}\kappa(r,\gamma)\varrho + \tau_{\varrho}(q,\gamma),
\end{equation}
\begin{equation}\label{eq:matt_eq_2}
        M(r) \ddot{r} = -C(r,\dot{r}) + N(r,\dot{r},\varrho) + \tau_{r}(r,\gamma),
\end{equation}
\begin{equation}\label{eq:matt_eq_3}
        \dot{\gamma} = -\xi_h \gamma , \quad\quad  \dot{\zeta} = -\xi_h (\zeta - A(r,\gamma)), 
\end{equation}
\end{thm}
where $\varrho$ are momenta (defined below), $\tau_{\varrho}(r,\gamma,\zeta)$ represents the conservative forces and moments resulting from gravity projected along the momenta directions.  The functions $\alpha(r,\gamma)$, $\beta(r,\gamma)$, and $\kappa(r,\gamma)$ are smoothly dependent upon the shape variables $r$ and advected parameter $\gamma$. $M(r)$ is the reduced mass-matrix, $C(r,\dot{r})$ contains the Coriolis terms, $N(r,\dot{r},\varrho)$ is the potential contribution, and $\tau_r(r,\gamma)$ represents the actuation forces associated with shape variables. $A(r,\gamma)$ is an invertible mass-like matrix which will arise in the reconstruction equation, defined later (\ref{eq:matt_eq_4}). The structure of (\ref{eq:matt_eq_1}) and  (\ref{eq:matt_eq_2}) follows from the reduced variational principle with
an extended base space consisting of the generalized momenta $\varrho = (\rho_1,\cdots,\rho_n)$ and shape-space variables $(r,\dot{r})$. (\ref{eq:matt_eq_1}) - (\ref{eq:matt_eq_3}) are the \textit{momentum equation}, \textit{shape dynamics}, \textit{advection equation}, and \textit{position dynamics} of the system, respectively. Together, they form a complete, reduced representation of the system dynamics. 

The momenta are defined as follows.  Recall the {\em group orbit} for the left action $\Phi_\mathtt{g}$(q) on a manifold $Q$ is the set $\mbox{Orb}(q) = \{\Phi_\mathtt{g}(q)|\forall \mathtt{g}\in G\}$. We denote the tangent space to the orbit at $q\in Q$ as $T_q\mbox{Orb}(q)$. When $G$ is the Lagrangian symmetry group, the tangent space $T_q\mbox{Orb}(q)$ represents
velocities along the symmetry directions. Let ${b_i}$ denote any basis of $T_q\mbox{Orb}(q)$.  For $SE(3)$, we choose the basis 
\small
\begin{equation} 
    T_q\mbox{Orb}(q) = \mbox{span}\left(\frac{\partial}{\partial s_{bx}},\frac{\partial}{\partial s_{by}},\frac{\partial}{\partial s_{bz}},\frac{\partial}{\partial \omega_{b_x}},\frac{\partial}{\partial \omega_{b_y}},\frac{\partial}{\partial \omega_{b_z}}   \right).
\end{equation}
\normalsize
The generalized momenta $\varrho_i$ are defined as $\varrho_i(\xi) = \left<b_i,\xi \right>$ along the symmetry directions. To recover the spatial motion of the system, we employ a {\em reconstruction equation} or {\em connection}, using trajectories in the reduced phase space. It defines a horizontal space of $T_qQ$ as $H_qQ\triangleq\mbox{Ker}(\mathcal{A}(q))$, where $\mathcal{A}$ is a principle connection form and describes motion along the fiber as the flow of a left-invariant vector field. See \cite{BKMM} and \cite{Ostrowski_1996}. The general form of connection is:
\small
\begin{equation} \label{eq:matt_eq_4}
     \xi_{\mathtt{h}} =\mathtt{h}^{-1}\dot{\mathtt{h}} = -A(r,\gamma) \dot{r} + B^{-1}(r,\gamma)\varrho,
\end{equation}
\normalsize
where $A(r,\gamma)$ and $B(r,\gamma)$ are mass and inertia tensors.



\subsection{Nonlinear Control Theory Review}
Consider steering problems in control-affine form: \small
\begin{equation} \label{eq:control_affine}
    \Sigma:\dot{q} = f(q) +
    \sum_{i=1}^{m} g_{i}(q)u_{i}
    \quad u \in \mathcal{U}\subset \mathbb{R}^{m},
\end{equation}
\normalsize
from an initial state $q_0\in Q$ to a final state $q_{f}\in Q$ by controls $u: [0,T]\to \mathcal{U}$. $f$ is the drift vector field. $g_{i}$, $i\in \{1,...,m\}$, are actuation vector fields which assumed to be real-valued smooth functions. Define a distribution $\Delta$ of the system $\Sigma$: $\Delta \triangleq \mbox{span}\{g_1,\cdots,g_m\}$. The followings are fundamentals for analyzing $\Sigma$'s controllability.

\textbf{Definition:} The {\bf reachable set}, denoted as $\mathcal{R}^V(q_0,\geq T)$, is the set of all points $q_{f}$ such that there exists an input $u(t)\in \mathcal{U}$, $0\leq t \leq T$, that steers the system (\ref{eq:control_affine}) from $q(0) = q_0$ to $q(T) = q_{f}$, where the trajectories $q(t)$, $0\leq t \leq T$, remain inside a neighborhood $V$ of $q_0\in Q$.

\textbf{Definition:} $\Sigma$ is {\bf Small-Time Locally Accessible} (STLA) from $q_0$ if the reachable set contains a full n-dimensional subset of $Q$ $\forall$ neighborhoods $V$ and all $ T>0$.

\textbf{Definition:} $\Sigma$ is {\bf Small-Time Locally Controllable} (STLC) from $q_0$ if the reachable set contains a neighborhood of $q_0$ $\forall$ neighborhoods $V$ and all $T>0$.

The \textbf{Lie bracket} (product) between vector fields $g_1(q)$ and $g_2(q)$ is
\small$
    [g_1(q),g_2(q)] \triangleq \frac{\partial g_1(q)}{\partial q}g_2(q) - \frac{\partial g_2(q)}{\partial q}g_1(q)$
\normalsize
, which quantifies how the derivative of vector field $g_2(q)$ varies along the flow of $g_1(q)$. The Lie bracket $[g_1(q),g_2(q)]$ (and higher order brackets) may enable infinitesimal movements locally in the system tangent space along directions that are not in $\Delta$. In summary, STLA is to access the Lie brackets' ability to generate independent actuation vector fields. STLC is to further examined the brackets' ability to overcome the drift $f$, leading to the following definition.

\textbf{Definition:} A Lie product is considered to be a {\bf bad bracket} if the drift term $f$ appears an odd number of times in the product and each control vector field $g_i,i\in \{1,...,m\}$, appears an even number of times (including zero). If a Lie product is not bad, it is a {\bf good bracket}.

\vskip 0.06 true in
\begin{thm} (\cite{HERMANN1963325}) \label{thm: larc}
The system (\ref{eq:control_affine}) is STLA from $q$ if it satisfies the Lie algebra rank condition (LARC):  $\overline{\mbox{Lie}}({f,g_1 ... , g_m})(q) = T_qQ$, where $\overline{\mbox{Lie}}(\cdot)$ is the closure of Lie algebra over the span of all input vector fields  and their iterated Lie brackets, $\{f(q),g_1(q),\cdots,g_m(q)\}$.
\end{thm}

For driftless systems, i.e. $f(q) = 0, \forall q\in Q$, STLA and STLC properties are equivalent, and LARC alone is sufficient to establish both properties. However, for systems with drift like an aerial manipulator robot, the following condition, proposed by \cite{Sussmann1987AGT}, is instead sufficient to conclude STLC.

\begin{thm} (\cite{Sussmann1987AGT}) \label{thm: sussmann}
System (\ref{eq:control_affine}) is STLC from $q^*$ if (1) $f(q^*) = 0$, (2) $\mathcal{U}$  is bilateral, (3) the LARC is satisfied by good Lie bracket terms up to degree $k$, and (4) any bad bracket of degree $j \leq k$ can be expressed as a linear combination of good brackets of degree less than $j$.
\end{thm}


\section{System Dynamics}\label{ses:sys_dyn}


Suppose $m_b$ and $I_b$ are the mass and inertia tensor of the bare multi-rotor structure, expressed in frame $B$. The kinetic energy of the this rigid body is  $K_b = \frac{1}{2}m_b\dot{s}_{eb}^{T}\dot{s}_{eb} + \frac{1}{2} \omega_{b}^T I_b\omega_{b}$
, and potential energy is $V_s = m_bge_3^{T}s_{eb}$, where vectors $e_1$, $e_2$, and $e_3$ be standard Cartesian basis vectors.

To properly capture the yaw dynamics and control, we separately model the rotors'  dynamics. 
Explicitly, the translational kinetic energy of the $j^{th}$ rotor $K_{t,t_j} = \frac{1}{2}m_{r}\dot{s}_{et_{j}}^{T}\dot{s}_{et_{j}}$ 
where $m_{r}$ denotes the mass of an individual rotor, and its rotational kinetic energy is
\small
\begin{equation} \label{eq:rotor_rotationalKE}
    K_{r,t_j} = \frac{1}{2}\left( J\Omega_j^2 + \omega_b^{T}I_r \omega_b + 2J\Omega_j e_{3}^T \omega_b \right),
\end{equation}
\normalsize
where \small$I_r \triangleq diag(a,a,J)$ \normalsize is the moment of inertia of the $j^{th}$ rotor w.r.t. frame \small$T_j$ \normalsize, and rotors are modeled as cylinders. 


Suppose the mass, length, and inertia tuple of the link 1 and 2  w.r.t. frame $L_1$ and $L_2$ are $\{m_1,d_1,I_1\}$ and $\{m_2,d_2,I_2\}$, respectively. The 2-link manipulator dynamics can be conveniently expressed using the manipulator Jacobian matrix:
\small
\begin{equation} 
    \left[ \begin{array}{cc}
         \dot{s}_{bl_1}  &
          \omega_{l_1}^b 
    \end{array} \right]^{T} =  \left[\begin{array}{cc}
         J_{l,bl_1}^b(\eta_1) &
         J_{r,bl_1}^b(\eta_1) 
    \end{array}\right]^T\dot{\eta}_1,
\end{equation}
\begin{equation} 
 \left[ \begin{array}{cc}
         \dot{s}_{bl_2}  &
          \omega_{l_2}^b 
    \end{array} \right]^{T} =  \left[\begin{array}{cc}
         J_{l,bl_2}^b(\eta_1,\eta_2) &        J_{r,bl_2}^b(\eta_1,\eta_2) 
    \end{array}\right]^T\left[\begin{array}{cc}
        \dot{\eta_1} &
        \dot{\eta_2}
    \end{array}\right]^{T},
\end{equation}
\normalsize
where $J_{bl_i}^{b} (\eta_1, \cdots ,\eta_i)\in\mathbb{R}^{6\times i}$ is the manipulator spatial Jacobian matrices defined in \cite{MurrayRobotManipulator}. Similar to the multi-rotor case, the kinetic and potential energy of each link can be calculated by summing the translational and rotational contributions. Details will be omitted.

\subsection{Overall System Dynamics}
Let $\dot{q} = [\dot{s}_b^{T},\omega_{b}^{T},\Omega_j,\dot{\eta}_1,\dot{\eta}_2]^{T}\in\mathbb{R}^{8+2n}$. The kinetic energy of the AM system can be rewritten as the following:
\small
\begin{equation*} \label{eq:mass_part}
    K_{AM}=\frac{1}{2}\dot{q}^{T}\mathcal{M}(\eta_1,\eta_2)\dot{q}= \frac{1}{2}  \dot{q}^{T}\left[\begin{array}{c|c|c|c}
          \mathcal{M}_{p} &\mathcal{M}_{p\omega} & 0_{3\times 2n} & \mathcal{M}_{pl} \\\hline
          \mathcal{M}_{p\omega}^{T} &\mathcal{M}_{\omega} & \mathcal{M}_{\omega r} & \mathcal{M}_{\omega l} \\\hline
          0_{2n\times 3} &\mathcal{M}_{\omega r}^{T} &  \mathcal{M}_{r} & \mathbf{0}_{2n\times 2} \\\hline
           \mathcal{M}_{pl}^{T} &\mathcal{M}_{\omega l}^{T} &\mathbf{0}_{2\times 2n}& \mathcal{M}_{l}\\
    \end{array} \right]\dot{q},
\end{equation*}
\normalsize
where $\mathcal{M}(\eta_1,\eta_2)\in \mathbb{R}^{(8+2n)\times(8+2n)}$  is the overall system mass matrix, which is block partitioned. $\mathcal{M}_p \in \mathbb{R}^{3\times 3}$, $\mathcal{M}_\omega \in \mathbb{R}^{3\times 3}$, $\mathcal{M}_r \in \mathbb{R}^{2n\times 2n}$,  and $\mathcal{M}_l \in \mathbb{R}^{2\times 2}$  are symmetric mass and inertia matrices of the multirotor structure, the rotors, and the manipulator with respect to $B$ frame. Matrices $\mathcal{M}_{p\omega} \in \mathbb{R}^{3\times 3}$, $\mathcal{M}_{pl} \in \mathbb{R}^{3\times 2}$, $\mathcal{M}_{\omega r} \in \mathbb{R}^{3\times 2n}$, and $\mathcal{M}_{\omega l} \in \mathbb{R}^{3\times 2}$  describe coupling effects. 
The total potential energy of the AM system with respect to inertial frame is 
 $V_{AM} = ge_3^T\left((m_b+2nm_{r}){s}_{eb} + m_1s_{el_1}+ m_{2}s_{el_2}\right).
$ 
Notice $V_{AM}$ is a function of the direction of gravity, the position of the vehicle, and manipulator joint angles, thus making the AM systems suitable to apply theorem \ref{thm:matt_thm}.


\subsection{Gravitational Potential Forces and Moments}
The previously defined advected parameter $\gamma$ and $\zeta$  satisfies the following advection equations:
\small
\begin{equation}\label{eq:advectionseq}
    \dot{\gamma} = -S(\omega_b)\gamma \quad \quad \dot{\zeta} = -S(\omega_b)\zeta + \dot{s}_b
\end{equation}
\normalsize
The Augmented Lagrangian, parametrized by $\gamma$ and $\zeta$,  is $SE(3)$-invariant by Theorem \ref{thm:matt_thm}.
The Augmented Lagrangian  $\overline{L}(q,\gamma,\zeta)$ for the system is:
\small
\begin{equation} \label{eq:overall_lag}
    \begin{array}{ll}
      \overline{L} = \frac{1}{2} \dot{q}^{T}\mathcal{M}(q)\dot{q}
    - g\gamma^T (m_t\zeta -m_{1}s_{bl_1}(\eta_1)- m_{2}s_{bl_2}(\eta_1,\eta_2)) 
    \end{array}
\end{equation}
\normalsize
where the total vehicle mass is lumped as $m_t = m_{b}+m_1+m_2+2nm_r$.
The conservative forces and momenta due to gravity can be derived from (\ref{eq:gravityreduction}):
\small
\begin{equation} \label{eq:dV}
\begin{array}{ll}
    -dV &= -\frac{\partial V}{\partial \gamma}\dot{\gamma}dt -\frac{\partial V}{\partial \zeta}\dot{\zeta}dt-\frac{\partial V}{\partial \eta_1}\dot{\eta_1}dt-\frac{\partial V}{\partial \eta_2}\dot{\eta_2}dt\\
\end{array}
\end{equation}
\normalsize
Explicitly, the force and torque of gravity $\tau_{\mathbf{p}}$ and $\tau_{\mathbf{l}}$ are:
\small
\begin{align}
    \tau_{\mathbf{p}} &= -gm_{t}\gamma^{T}, \\
    \tau_{\mathbf{l}} & =g\left(m_1 s_{bl_1}^{T} + m_2s_{bl_2}^{T}\right)S(\gamma)\label{eq:taul_pt1}
    \\&-g\gamma^{T}\left(m_1\frac{\partial s_{bl_1}}{\partial \eta_1} + m_2   \left[ \begin{array}{cc}
 \frac{\partial s_{bl_2}}{\partial \eta_1} &  \frac{\partial s_{bl_2}}{\partial \eta_2}\label{eq:taul_pt2}
    \end{array}\right] \mathbf{1}_{2\times1}\right)e_2.
\end{align}
\normalsize

\subsection{Non-Conservative Forces and Moments}

This section describes the non-conservative forces arising from aerodynamic effects and manipulator actuation, and we only address aerodynamic effects near hovering conditions. The thrust force produced by the $j^{th}$ rotor is modeled as the product of a lumped thrust coefficient $c_{T}$ with the rotating velocity squared. Letting \small $\mathbf{\Omega} = \left[\begin{array}{ccc}
         \Omega_{1}  &
         \cdots &
         \Omega_{2n}   
    \end{array} \right]^{T}$\normalsize, be the rotor RPMs, and
    the total thrust is $\tau_{f} =c_{T}\mathbf{\Omega}^{T}\mathbf{\Omega}$.
The torques imparted on the vehicle in the frame $B$ are:
    \small
 \begin{eqnarray} 
    \left[\begin{array}{c}
         \tau_{r}  \\
         \tau_{p}  \\
         \tau_{y}
    \end{array} \right] &=& \left[\begin{array}{cccc}
    0 & rc_{T}&0 &\cdots \\
    -rc_{T}&0 & rc_{T} &\cdots \\
    -c_{Q} & c_{Q}& -c_{Q} & \cdots 
    \end{array} \right]
    \left[\begin{array}{c}
         \Omega_{1}^2   \\
         \vdots \\
         \Omega_{2n}^2 
    \end{array}
    \right]
      \label{eq:thrusttorque_b} \triangleq W\left[\begin{array}{c}
         \Omega_{1}^2   \\
         \vdots \\
         \Omega_{2n}^2 
    \end{array}
    \right]
\end{eqnarray} 
\normalsize
where $c_{Q}$ is the lumped resistive torque coefficient that depends on rotor geometry. Together, they form the non-conservative forces and moments vector $\tau_{nc} \triangleq   \left[\begin{array}{ccc}
         \tau_{r} & 
         \tau_{p} &
         \tau_{y}
    \end{array} \right]^T $. The manipulator joint actuators are modeled as a bilateral torque input at the joints. Explicitly, the torque input at the revolute joint connecting aerial base and link 1 is $\tau_{l_1}$ and connecting link 1 and link 2 is $\tau_{l_2}$. 


Variable RPM rotor thrusts are unilateral, making it incompatible with the second requisite for Sussmann's theorem. 
Alternatively, we adopt the time rate of change of rotor RPMs as control inputs. It is physically reasonable to assume that $\{\dot{\Omega}_1, \cdots,\dot{\Omega}_n\}$ can be driven bilaterally since the rotor dynamics can be modeled as $ \dot{\Omega}_i = \frac{1}{J}(\tau_{d}+\tau_r)$, where $\tau_d>0$ is the motor drive torque and $\tau_{r}<0$ is a passive resistive torque due to air damping. For $\dot{\Omega}_i = 0$, $\tau_{d}+\tau_{r}$ implies balance between the drive and resistive torque, and $\tau_{r}<0$ when the rotor is spinning. From this equilibrium position, one can increase $\tau_d$ to produce positive $\dot{\Omega}_i$. Vice versa, if $\tau_d$ decreases yielding $\tau_d + \tau_r<0$, a negative rate of change in RPM is possible, $\dot{\Omega}_i <0$. 


\subsection{Reduced Dynamics:}
In the standard basis for $\mathfrak{se}(3)$, the generalized rotation momentum and translational momentum in $B$ frame are defined as $\mathbf{l} \triangleq \frac{\partial \overline{l}}{\partial \omega_b}\in T_q^*SO(3)$ and  $\mathbf{p}\triangleq \frac{\partial \overline{l}}{\partial \dot{s}_{b}}\in T_q^*\mathbb{R}^{3}$:
\small
\begin{equation*} \label{eq:momenta}
    \left[\begin{array}{c}
         \mathbf{p}  \\
         \mathbf{l}
    \end{array} \right] =     
    \underbrace{\left[\begin{array}{cc}
         \mathcal{M}_{p} & \mathcal{M}_{p\omega}    \\
        \mathcal{M}_{p\omega}^T & \mathcal{M}_{\omega}   
    \end{array}\right]}_\text{$\triangleq \mathcal{M}_s$}   \left[\begin{array}{c}
         \dot{s}_{b}  \\
         \omega_b
    \end{array} \right]+  \left[\begin{array}{cc}
         \mathcal{M}_{sr} &\mathcal{M}_{sl}
    \end{array} \right]     \left[\begin{array}{c}
        \mbox{diag}(\mathbf{\Omega})\mathbf{\Omega}\\
         \dot{\eta}_1  \\
         \dot{\eta}_2 
    \end{array} \right].
\end{equation*}
\normalsize
where \small$\mathcal{M}_{sr} \triangleq [ 0_{3\times 2n}, \mathcal{M}_{\omega r}]^{T}$ \normalsize  and \small $\mathcal{M}_{sl} \triangleq [ \mathcal{M}_{pl}, \mathcal{M}_{\omega l}]^{T}.$ \normalsize Since the AM system is constraint free, the connection is simply:
\small
\begin{equation} \label{eq:connectioneq}
\left[\begin{array}{c}
         \dot{s}_{b}  \\
         \omega_b
\end{array}\right]=\mathcal{M}_s^{-1}\left(\left[\begin{array}{c}
         \mathbf{p}  \\
         \mathbf{l}
\end{array} \right]-\left[\begin{array}{cc}
         \mathcal{M}_{sr}\mbox{diag}(\mathbf{\Omega})&\mathcal{M}_{sl}\\
\end{array} \right]\left[\begin{array}{c}
        \mathbf{\Omega}\\
         \dot{\eta}_1  \\
         \dot{\eta}_2
\end{array} \right]\right).
\end{equation}
\normalsize
Following \cite{OSTROWSKI1998185}'s work, the non-conservative/ dissipative forces $\tau_{f}$ and torques \eqref{eq:thrusttorque_b} can be projected along the acting momentum directions. Since the symmetry directions are a simple basis of $\mathfrak{se}$(3), the momentum equation including conservative forces $-dV$ and nonconservative forces and moments $\tau_{nc}$ becomes:
\small
\begin{equation} \label{eq:generalizedmomentumode}
    \dot{\varrho}_i = \left<\frac{\partial \overline{l}}{\partial \xi},[\xi,b_i] + \left< \frac{\partial b_i}{\partial r},\dot{r} \right>\right> +\left<-dV,b_i \right> +\left<\tau_{nc},b_i\right>
\end{equation}
\normalsize
Substituting $\tau_{f}$, (\ref{eq:dV}), and (\ref{eq:thrusttorque_b}), we can obtain the translational and rotational momenta differential equations:
\small
\begin{equation}\label{eq:momenta}
\begin{array}{ll}
    \dot{ \mathbf{p}} &= \mathbf{p} \times \omega_b + \tau_{\mathbf{p}} +  c_{T}\mathbf{\Omega}^{T}\mathbf{\Omega}e_3 \\
    \dot{ \mathbf{l}} &= \mathbf{p} \times \dot{s}_{b} + \mathbf{l} \times \omega_b+\tau_{\mathbf{l}} + W \mbox{diag}(\mathbf{\Omega})
      \mathbf{\Omega}  \end{array}
\end{equation}
\normalsize
Next, we derive the dynamics of the manipulator. Let $\eta = [\eta_1, \eta_2]^{T}$. With system Lagrangian (\ref{eq:overall_lag}), one can apply the Euler-Lagrange equation for the shape variable $\eta_i$ to extract the following the second-order ODE:
\small
\begin{equation} \label{eq:manipulator_dynamics}
\ddot{\eta} = \mathcal{M}_{\eta}^{-1}\left(-f_{\eta} + g_{\eta}\dot{\mathbf{\Omega}} +\left[\begin{array}{cc}
         \tau_{l_1} &
         \tau_{l_2}
    \end{array} \right]^{T}\right).
\end{equation}
\normalsize
where \small $\mathcal{M}_{\eta} = \mathcal{M}_l+\mathcal{M}_{sl}^{T}\mathcal{M}_s^{-1}\mathcal{M}_{sl}$\normalsize. Further, $f_{\eta}$ and $g_{\eta}$ are:
\small
\begin{align*}
    f_{\eta} &= \dot{\mathcal{M}}_l\dot{\eta} + \frac{d}{dt}\left(\left(\mathcal{M}_{sl}^{T}\mathcal{M}_{s}^{-1} \right)\left(\left[\begin{array}{c}
         \mathbf{p}  \\
         \mathbf{l}
\end{array} \right]-\left[\begin{array}{cc}
         \mathcal{M}_{sr}\mbox{diag}(\mathbf{\Omega})&\mathcal{M}_{sl}\\
\end{array} \right]\left[\begin{array}{c}
        \mathbf{\Omega}\\
         \dot{\eta}
\end{array} \right]\right)\right) \\
&+ \frac{1}{2}\frac{\partial (\dot{q}^{T}\mathcal{M}(q))}{\partial \eta}\dot{q} -g\left(m_{1}\frac{\partial s_{bl_1}}{\partial \eta} +m_{2}\frac{\partial s_{bl_2}}{\partial \eta}\right)^{T}\gamma,\\
g_{\eta} &= -\mathcal{M}_{sl}^{T}\mathcal{M}_{s}^{-1}\mathcal{M}_{sr}\mbox{diag}(\mathbf{\Omega}).
\end{align*}
\normalsize

In summary, the reduced EOM for AM system are: the momenta equations \eqref{eq:momenta}, the shape dynamics \eqref{eq:manipulator_dynamics}, and the advection equations \eqref{eq:advectionseq}. The Connection (\ref{eq:connectioneq}) can be incorporated to reconstruct the system's motion in $SE(3)$.

\section{Nonlinear Controllability Assessment}
\label{ses:nonlinearC}


To prepare for nonlinear controllability analysis, the EOM need to be organized into control affine form. We defined the overall system state variable to be:
\small
\begin{equation*}
    \mathbf{q} \triangleq [\mathbf{p}^{T},\mathbf{l}^{T},\phi,\theta,\psi,\eta_1,\eta_2,\dot{\eta}_1,\dot{\eta}_2,\Omega_1,\cdots,\Omega_{2n}]^{T}\in \mathbb{R}^{2n+13}
\end{equation*}
\normalsize
where we use the Euler angle instead of the advected parameter $\gamma$ to track the the multirotor orientation.
Further, the control affine form of the full AM system is:
 \small
\begin{equation}  \label{eq:control_affine_form_final}
    \dot{\mathbf{q}} =  f(\mathbf{q}) + G(\mathbf{q})\mathbf{u},
\end{equation}
\normalsize
where the control input of the system is 
\small
\begin{equation}
    \mathbf{u} = [\tau_{l_1},\tau_{l_2},\dot{\Omega}_1,\dot{\Omega}_2, \cdots, \dot{\Omega}_{2n}]^{T}\in \mathbb{R}^{2n+2}.
\end{equation}
\normalsize
Further, the drift term and actuation vector fields are
\small
\begin{equation*}
    f(\mathbf{q}) = \left[\begin{array}{c}
    \mathbf{p}\times\omega_{b}- m_{t}g\gamma+e_3c_{T}\mathbf{\Omega}^{T}\mathbf{\Omega}\\
   \mathbf{p} \times \dot{s}_{b} + \mathbf{l} \times \omega_b+\tau_{\mathbf{l}} + W \mbox{diag}(\mathbf{\Omega})
      \mathbf{\Omega}\\
     T(\mathbf{\Theta})^{-1}\omega_{b}\\
    \dot{\eta}\\
      -\mathcal{M}_{\eta}^{-1}f_{\eta} \\
         0_{(2n)\times (2n+2})
    \end{array} \right], 
\end{equation*}
\normalsize
\small
\begin{equation*}
G(\mathbf{q}) = \left[\begin{array}{ccc}
      g_1(\mathbf{q})  & \cdots & g_{2n+2}(\mathbf{q})
    \end{array} \right]= \left[\begin{array}{cc}
         0_{11\times 2} & 0_{11\times 2n}\\
         \mathcal{M}_{\eta}^{-1} & g_\eta \\
         0_{2n\times 2} & I_{2n\times 2n}\\
    \end{array} \right], 
\end{equation*}
\normalsize
respectively, where unknown states $\omega_{b}$ and $\dot{s}_b$ can be expressed in terms of $\mathbf{q}$ using the connection (\ref{eq:connectioneq}) and $\tau_{\mathbf{l}}$ is a function of $\gamma$ and $\eta$ given in (\ref{eq:taul_pt1}) - (\ref{eq:taul_pt2}).


\subsection{Equilibrium Condition}
Nonlinear controllability can only be assessed in the neighborhood of equilibrium conditions.
Setting  $\dot{\mathbf{q}} = 0$, one can verify the following relationship must be true:
\small
\begin{align}
    0_{1\times 9}& = \left[\begin{array}{ccccccccc}
         \mathbf{p}_x^* & \mathbf{p}_y^* & \mathbf{p}_z^*& \mathbf{l}_x^*& \mathbf{l}_y^* &\phi^* &\theta^* & \dot{\eta}_1^* & \dot{\eta}_2^*
    \end{array} \right], \label{eq:eqlib_c1}\\
    \mathbf{l}^*_z &= J(\mathbf{\Omega}^*)^{T}\mathbf{\Omega}^*\label{eq:eqlib_c2}, \quad \quad     c_{T}(\mathbf{\Omega}^*)^{T}\mathbf{\Omega}^* = m_{t}g,\\
    W\mbox{diag}(\mathbf{\Omega}^*)\mathbf{\Omega}^* &= -ge_3 \times\left(m_1 s_{bl_1}(\eta_1) +  m_2 s_{bl_2}(\eta_1,\eta_2)\right),\label{eq:eqlib_c3}\\&+ ge_{3}^{T}\left(m_1\frac{\partial s_{bl_1}}{\partial \eta_1} + m_2
    \left[ \begin{array}{cc}
 \frac{\partial s_{bl_2}}{\partial \eta_1} &  \frac{\partial s_{bl_2}}{\partial \eta_2}
    \end{array}\right] \mathbf{1}_{2\times1}\right)e_2.\nonumber
\end{align} 
\normalsize
Further, $\eta_1$, $\eta_2$, and $\psi$ are free, meaning any combinations of the free states can be an equilibrium condition of the overall system. Physically speaking, these conditions correspond to a hovering AM at arbitrary vehicle position and manipulator configurations with zero roll and pitch. 

\subsection{Small-Time Locally Accessible and Controllability}

\begin{thm}
All AMs are Small-Time Locally Accessible (STLA) evaluated at the equilibrium conditions (\ref{eq:eqlib_c1}) - (\ref{eq:eqlib_c3}), except at manipulator singularity $\eta= [\pi/2,0]^{T}$.
\begin{pf}  
By Theorem \ref{thm: larc}, we must check that system (\ref{eq:control_affine_form_final}) satisfied the LARC when evaluated at the equilibrium condition, equivalently $\overline{\mbox{Lie}}({f,g_1 ... , g_m})(\mathbf{q}^*) = T_{\mathbf{q}^*}Q$.

First note that $g_{1}(\mathbf{q}^*)$ and $g_{2}(\mathbf{q}^*)$ are functions of the free state $\eta_2$, and $\forall \eta_2$, $\mbox{rank}([g_{1}(\mathbf{q}^*), g_{2}(\mathbf{q}^*)]) = 2$, spanning the states corresponds to $\ddot{\eta}_1$ and  $\ddot{\eta}_2$. Despite coupling with the $\ddot{\eta}_1$ and  $\ddot{\eta}_2$, the remaining first degree actuation fields have $\mbox{rank}(g_{3},\cdots,g_{2n+2}) = 2n$, annihilating any nonzero $\dot{\mathbf{\Omega}}$ with appropriate combinations of $\mathbf{u}$. Since $g_1$ and $g_2$ corresponds to manipulator maneuvers in the $x_{b}-z_{b}$ plane, the Lie brackets between $g_{1}$, $g_{2}$ with the drift $f$ are:
\small
\begin{equation*}
\begin{array}{ll}
    \left[f,g_{i}\right]\big|_{\mathbf{q}^*}  = \left[\begin{array}{ccccccccc}
        0^{\dot{\mathbf{p}}}_{1\times 3} & f^{[\dot{\mathbf{l}}_{x},\dot{\mathbf{l}}_{y}]}_{1\times 2} &0^{\dot{\mathbf{l}}_{z}} &f^{\dot{\mathbf{\Theta}}}_{1\times3} &  f^{[\dot{\eta},\ddot{\eta}]}_{1\times 4} & 0^{\mathbf{\dot{\Omega}}}_{1\times 2n}
    \end{array}\right]^{T},
\end{array}
\end{equation*}
\normalsize
which are two independent actuation fields that primarily influences $\dot{\eta}_1$ and $\dot{\eta}_2$ and also perturbs the angular velocities and rotational momentum because of the shifting in center of gravity. The remaining $2n$ actuation vector fields together span $\dot{\mathbf{l}}$ and $\dot{\mathbf{p}}_z$. For $i\in\{3,2n+2\}$, we have
\small
\begin{equation*}
    \left[f,g_{i}\right]\big|_{\mathbf{q}^*}  = \left[\begin{array}{ccccccccc}
       0^{[\dot{\mathbf{p}}_x,\dot{\mathbf{p}}_y]}_{1\times2} & f^{\dot{\mathbf{p}}_z} & f^{\dot{\mathbf{l}}}_{1\times 3}  &f^{\dot{\mathbf{\Theta}}}_{1\times3}  & f^{[\dot{\eta},\ddot{\eta}]}_{1\times 4} & 0^{\mathbf{\dot{\Omega}}}_{1\times 2n}
    \end{array}\right]^{T},
\end{equation*}
\normalsize
where $f^{\dot{\mathbf{\Theta}}}_{1\times 3}$ is non-zero because of the unbalance in rotor thrust leads to a non-zero resistive torque resulting in non-zero body rates. The manipulator joints are also coupled with the rotor thrust rate via $g_\eta$ in (\ref{eq:manipulator_dynamics}). Therefore, any perturbation in rotor thrust will leads to a non-zero rotation. Aided by the asymmetry, all degree two brackets together add six more independent control distributions. Looking at the third degree brackets and for $i\in\{1,2\}$,
\small
\begin{equation*}
    \left[[f,g_{i}],f\right]\big|_{\mathbf{q}^*}  = \left[\begin{array}{ccccccc}
          f^{[\dot{\mathbf{p}}_x,\dot{\mathbf{p}}_y]}_{1\times2} & 0^{\dot{\mathbf{p}}_z}  & f^{\dot{\mathbf{l}}}_{1\times 3} &f^{\dot{\mathbf{\Theta}}}_{1\times 3} & f^{[\dot{\eta},\ddot{\eta}]}_{1\times 4}& 0^{\mathbf{\dot{\Omega}}}_{1\times 2n}
    \end{array}\right]^{T},
\end{equation*}
\normalsize
which provides two more independent bases for $T_{\mathbf{q^*}}Q$. Physically speaking, the third degree brackets with $g_1$ and $g_2$ are similar to the second degree brackets where shifting the vehicle center of gravity perturbs the hovering motion. It is important to note that the perturbation will come in the form of forces which leads to nonzero translational accelerations. Further, the remaining third order brackets provided 3 more independent bases. For $i\in\{3,2n+2\}$,
\small
\begin{equation*}
    \left[[f,g_{i}],f\right]\big|_{\mathbf{q}^*}
    = \left[\begin{array}{ccccccccc}
   f^{[\dot{\mathbf{p}}_x,\dot{\mathbf{p}}_y]} &0^{\dot{\mathbf{p}}_z} & f^{\dot{\mathbf{l}}}_{1\times 3}& f^{\dot{\mathbf{\Theta}}}_{1\times 3} &
  & f^{[\dot{\eta},\ddot{\eta}]}_{1\times 4}&
  0^{[\mathbf{\dot{\Omega}}]}_{1\times 2n}
    \end{array}\right]^{T}.
\end{equation*}
\normalsize
Together, there are $13+2n$ good brackets that can be used in spanning the tangent space $T_{\mathbf{q}^*Q}$ except at the manipulator singularity ($\eta = [\pi/2,0]$). Therefore, by theorem \ref{thm: larc}, we conclude AM are STLA evaluated at the equilibrium condition (\ref{eq:eqlib_c1})-(\ref{eq:eqlib_c3}) except at system singularity. $\blacksquare$ 

\normalsize
\end{pf}

\end{thm}


\begin{thm}
All AMs are Small-Time Locally Controllable (STLC) evaluated at the equilibrium conditions (\ref{eq:eqlib_c1}) - (\ref{eq:eqlib_c3}), except at manipulator singularity $\eta= [\pi/2,0]^{T}$.
\begin{pf}  
By definition of good and bad brackets, all Lie brackets used to establish STLA are good brackets, and the highest degree bracket used in spanning $T_{\mathbf{q}^{*}}Q$ is \textbf{three}. To apply theorem \ref{thm: sussmann}, we need to show all bad brackets of degree $j\leq 3$ can be expressed as a linear combination of good brackets of degree less than $j$. By its definition, there will be no bad brackets associated with an even degree. The third degree bad brackets can be expressed generally as $[[f,g_i],g_i],\forall i\in \{1,\cdots,2n+2\}$. More specifically, for $i \in \{1,2\}$ the bad brackets evaluated at the equilibrium condition takes the following form:
\small
\begin{equation*}
     [[f,g_{i}],g_i]\big|_{\mathbf{q}^*} =\left[\begin{array}{cccc} 
         0^{[\dot{\mathbf{p}},\dot{\mathbf{l}},\dot{\mathbf{\Theta}}]}_{1\times 12}& 0^{\dot{\eta}}_{1\times 2}&  f^{\ddot{\eta}}_{1\times 2}&         0^{\mathbf{\dot{\Omega}}}_{1\times 2n}
     \end{array} \right],
\end{equation*}
\normalsize
which can be annihilated with good, degree one bracket $g_1$ and $g_2$. The remaining $2n$ third degree bad brackets take the following form where $i\in \{3,\cdots,2n\}$:
\small
\begin{equation*}
   \left[\begin{array}{cccccc}   [[f,g_{i}],g_i]\big|_{\mathbf{q}^*} =0^{[\dot{\mathbf{p}}_x,\dot{\mathbf{p}}_y]}_{1\times2} & f^{\dot{\mathbf{p}}_z} & f^{\dot{\mathbf{l}}}_{3\times1} &f^{\dot{\mathbf{\Theta}}}_{3\times 1} & f^{[\dot{\eta},\ddot{\eta}]}_{4 \times 1} &0^{\dot{\mathbf{\Omega}}}_{2n \times 1}
    \end{array}\right],
\end{equation*} 
\normalsize
which can be annihilated by good degree two brackets $[f,g_i]$ and degree one brackets $g_1$ and $g_2$.
The final bad bracket is the drift term evaluated at equilibrium conditions. Since by the definition of equilibrium condition, we have $\dot{\mathbf{q}}^* =  f(\mathbf{q}^*) + G(\mathbf{q}^*)\mathbf{u^*}= 0$. 
Thus, all bad brackets of degree $k\leq 3$ can be written as combinations of good Lie brackets of degree $j\leq k$, establishing STLC.
$\blacksquare$
\end{pf}
\end{thm}

It is important to note that when interpreting these results, the net motions generate from higher degree Lie brackets ($d>1$) are ``slower" than motions driven directly by actuation vectors fields $g_1,\cdots,g_{m}$. In fact, the net motions are $O(\epsilon^{d})$ for time $O(\epsilon)$, where $\epsilon \ll 1$. See \cite{Choset_book}. Aided by net external motions resulting from the coupling of internal shape changes, the AM system can achieve local accessibility and controllability.  However, we suggest the more valuable takeaway from our SLTA and SLTC analysis should be the compositions of good and bad lie brackets since they contain information about how well the system can maneuver.

In summary, Lagrangian reduction process, directly produces the minimum set of first-order dynamical equations, substantially simplifies AM systems' controllability analysis. Using the connection, the ``reduced'' EOM can be reformulated into the control-affine form effortlessly. In comparison, to assess controllability without the reduced dynamics, using the EOM given by \cite{planararm_ex1} requires a costly symbolic inversion of a $(12+2n)\times (12+2n)$ sized mass matrix to reformulate. Further, the reduction and reconstruction process can be generalized to analyze other flying vehicles that include conventional quadcopter, tilt-rotors, multi-rotors with asymmetric rotor placements, tethers, and pendulum appendages. Concurrently, we are extending the geometric reduction and controllability analysis to more general flying vehicles.




\section{Conclusion}\label{ses:Conclusion}

This paper analyzed two aspects of a class of aerial-manipulator robots.  First, aerial manipulator robots' EOM was developed using Lagrangian reduction and reconstruction despite having broken symmetry.  Second, the AM system properties, STLA and STLC, are formally analyzed. We have concluded that underactuated multi-rotor with planar two-link manipulators are STLA and STLC near equilibrium condition. A physical interpretation of this result is if the AM is hovering, there exists control actions to maintain hover while the manipulator joints track a non-singular and smooth trajectory. However, our STLC result should be applied with caution since a controllability certificate does not assess the relative control effort needed to realize the trajectory or hovering state. 


\bibliography{ifacconf} \normalsize             
                                                   








\end{document}